\documentclass[twocolumn,preprint]{revtex4}
\usepackage{epsfig}
\usepackage{graphicx}
\usepackage{slashed}
\usepackage{xcolor}

\usepackage{bm}
\usepackage{subfigure}
\usepackage{amsmath} 
\usepackage{amssymb}

\newcommand{\rd}{\textrm{d}}
\begin{document}
%
\title{Initial state radiation in $e^+e^-$ annihilation }
\date{\today}
\author{G\"oran F\"aldt}\email{goran.faldt@physics.uu.se}  
\affiliation{ Department of physics and astronomy, 
Uppsala University,
 Box 516, S-751 20 Uppsala,Sweden }

\begin{abstract}
In two earlier papers it was demonstrated that  Lorentzian and Galilean symmetries could both be useful in 
the analysis of the annihilation reaction  
$e^+ e^- \rightarrow \gamma \Lambda(\rightarrow p\pi^-) \bar{\Lambda}(\rightarrow \bar{p}\pi^+)$. It was also demonstrated that any pair 
of hyperon form factors would be acceptable, but that the $\{G_E, G_M\}$ pair 
would probably be the simplest one to handle.

\end{abstract}
\maketitle
%
%
%

\newpage
\section{Introduction}\label{ett}

The {\slshape BABAR} Collaboration \cite{BaBar} has  measured initial-state-radiation in
the annihilation channel,   \\
$e^+ e^- \rightarrow \gamma \Lambda(\rightarrow p\pi^-) \bar{\Lambda}(\rightarrow \bar{p}\pi^+)$. \\
Theoretical analyses of the same reaction are presented in 
 ref.\cite{Novo}, for the $\Lambda\bar{\Lambda}\gamma$ final state 
with single hyperon polarization, 
and in ref.\cite{Czyz}, for the $\Lambda\bar{\Lambda}\gamma$ final state 
with double  hyperon polarizations. 
A Lorentz-covariant  description  of the cross-section-distribution 
functions, including those representing the hyperon decays, but with sums
over polarizations, is  presented in
 ref.\cite{Ent1}.

In the present investigation we work in the covariant Galilean  
formalism  \cite{EPJ58}, 
and choose as form factors the $\{G_E, G_M\}$ pair. 
 Our aim is to demonstrate how, in this formalism, the phase-space
density can be treated.  

\begin{figure}[ht] 
\scalebox{0.60} {  \includegraphics{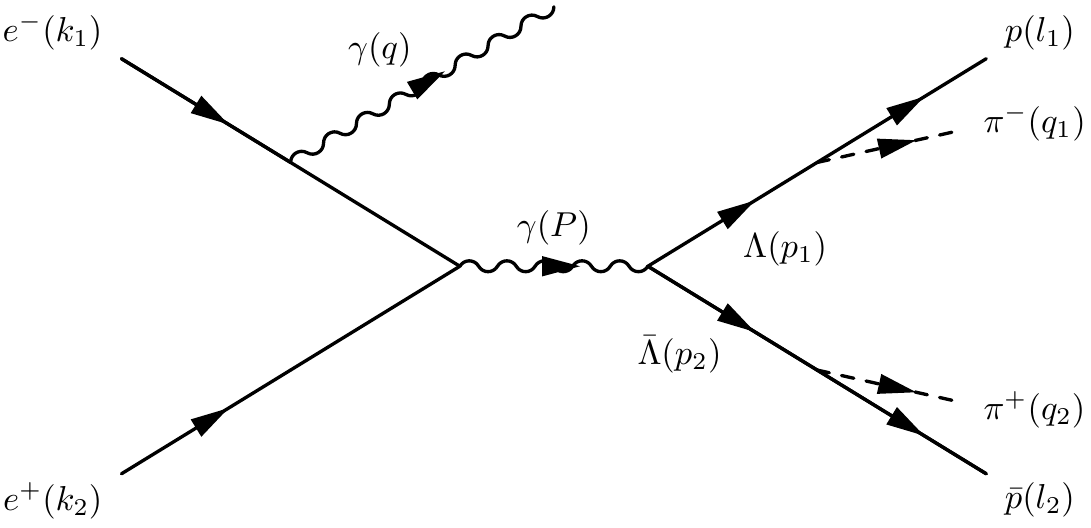} }\\ 
\vspace{7mm}
\scalebox{0.60} { \includegraphics{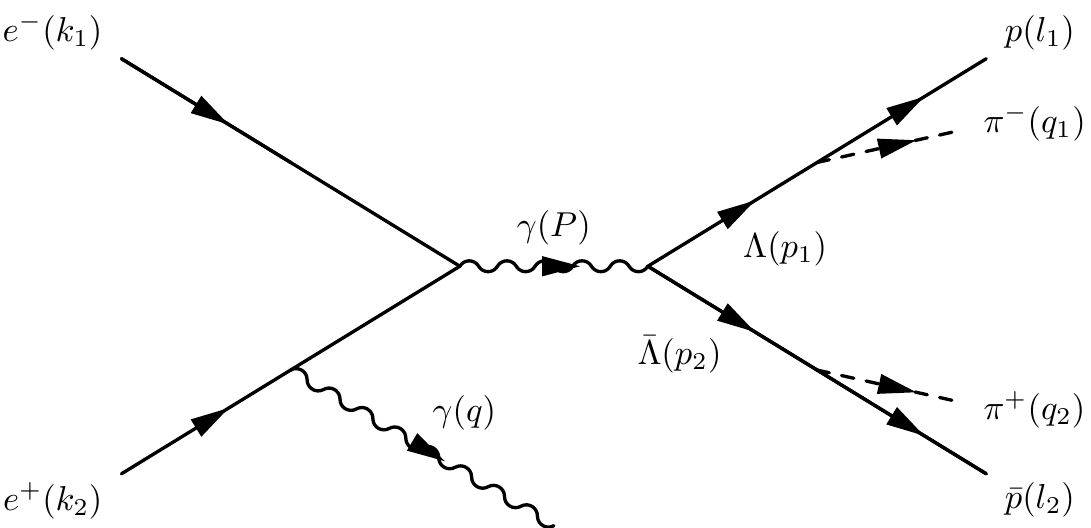}  }
\caption{Graphs included in our calculation of the reaction 
$e^+ e^- \rightarrow \gamma \Lambda(\rightarrow p\pi^-) \bar{\Lambda}(\rightarrow \bar{p}\pi^+)$.}
\label{F1-fig}
\end{figure}

%
%
\newpage
\section{Cross section}\label{tre}

 The cross-section distribution  for the 
initial-state-radiation reaction under consideration, is factorized 
as  in refs.\cite{Ent1,EPJ58}, 
\begin{eqnarray}
	\rd \sigma&=& \frac{1}{2\sqrt{\lambda(s,m_e^2,m_e^2)}}
	\,{\cal{K}}\,\overline{|{\cal{M}}_{red}|^2}\nonumber \\ &\times&
	   \textrm{dLips}(k_1+k_2;q,l_1,l_2,q_1,q_2),	
		 \label{dsigma}
\end{eqnarray}	  
where the average over the squared matrix element indicates summation over final-state nucleon spins and average over initial-state lepton spins,  and 
 with the ${\cal{K}}$ factor
 \begin{eqnarray}
	{\cal{K}}= \frac{(4\pi\alpha)^3}{(P^2)^2}    & \cdot &   
	\frac{1}{(s_1 -M_\Lambda^2)^2 +M_\Lambda^2 \Gamma_\Lambda^2 } 
	\nonumber	
	\\  &\cdot &	\frac{1}{(s_2 -M_\Lambda^2)^2+M_\Lambda^2 \Gamma_\Lambda^2}	
. \label{KM-factor}
\end{eqnarray} 
Notations for variables and parameters are all explained 
in FIG.1 of refs.\cite{Ent1,EPJ58}. In particular, $s_1=(l_1+q_1)^2$ 
and $s_2=(l_2+q_2)^2.$
A consequence of the factorization of the cross section is 
the factorization of the squared matrix element, 
\begin{equation}
	\overline{|{\cal{M}}|^2}={\cal{K}}\overline{|{\cal{M}}_{red}|^2},\label{Msq_andK}
\end{equation}
with ${\cal{K}}$ of eq.(\ref{KM-factor}).

%
\newpage
\section{Reduced matrix element}\label{sec3}

The cross-section distribution, or rather the covariant square of the annihilation 
matrix element $\overline{|{\cal{M}}_{red}|^2}$, 
 is obtained by contracting hadronic $H_{\mu\nu}$
 and leptonic $L^{\mu\nu}$ tensors, so that
\begin{equation}
	\overline{|{\cal{M}}_{red}|^2}=L^{\mu\nu}H _{\mu\nu}.
\end{equation}

The right-hand-side of this equation can be rewritten as 
 a sum of four terms,
\begin{eqnarray}
	\overline{|{\cal{M}}_{red}|^2} &=&\bar{R}_\Lambda R_\Lambda M^{RR}+\bar{R}_\Lambda S_\Lambda M^{RS} \nonumber \\
	   &+&\bar{S}_\Lambda R_\Lambda M^{SR} +\bar{S}_\Lambda S_\Lambda M ^{SS},\label{MM-decomp}
\end{eqnarray}
with  coefficients  $R_\Lambda ,S_\Lambda$ and 
$R_{\bar{\Lambda}},S_{\bar{\Lambda}}$ 
that 
refer  to the $\Lambda$ and $\bar{\Lambda}$ decay constants 
  of ref.\cite{Ent1},
and with $R$ the spin-independent  and $S$ the spin-dependent ones. 

From the structure of the lepton tensor,  eq.(24) of ref.\cite{Ent1}, it follows that 
each of the $M^{XY}$ functions of eq.(\ref{MM-decomp})  has two parts, 
one $A$-part and one $B$- part, 
\begin{equation}
	M^{XY}=-a_y A^{XY}(G_M,G_E) -b_yB^{XY}  (G_M,G_E) ,\label{MXY}
\end{equation}
where the $A^{XY}$ factor is obtained by contracting the hadron tensor 
with the symmetric tensor
$	k_{1\mu}k_{1\nu} + k_{2\mu}k_{2\nu}$, 
and the $B^{XY}$ factor by contraction with the tensor $ g_{\mu\nu}.$ 
For details see ref.\cite{Ent1}. The weight factors 
$a_y$ and $b_y$ are defined in 
the appendix.

 The arguments of the form factors can be chosen equal to $P^2$. In particular, when 
$P^2=4M^2$ then $G_M=G_E$.

The functions $A^{XY}$ and $B^{XY}$ are bilinear forms of $G_M$ and 
$G_E$, and expanded accordingly, in which case 
\begin{eqnarray}
A^{XY}(G_M,G_E) &=&|G_M|^2 {\cal{L}}^{AXY}_1 + |G_E|^2 {\cal{L}}^{AXY}_2
\nonumber \\
	  &+& 2 \Re (G_MG_E^{\star}) {\cal{L}}^{AXY}_3
		\nonumber \\ & + &2\Im (G_MG_E^{\star}) {\cal{L}}^{AXY}_4,   
				\label{CDdef}
\end{eqnarray}
and similarly for $B^{XY}$. We refer to the set of  functions 
\{${\cal{L}}$\} as co-factors.

\newpage

 \section{The co-factors}   

 The most remarkable fact 
	about the new set is that several  co-factors vanish;
	\begin{equation}
	{\cal{L}}_3^{ARR}={\cal{L}}_3^{BRR}=0. 
	\end{equation}
Also, as we shall see,  ${\cal{L}}_3^{BSS}= 0$
	but ${\cal{L}}_3^{ASS}\neq 0.$ Our results are the following.
	
		Co-factors suffixed ${ARR}$;
	\begin{eqnarray}
		{\cal{L}}_1^{ARR} &=&(2 \epsilon \omega)^2\,  \bigg[ \Phi_{\bot}
		 \bigg], \\
				{\cal{L}}_2^{ARR}& =&(2 \epsilon \omega)^2 \,
				\bigg[ \Phi_f \,  \frac{1}{ \gamma_{\Lambda}^2}\bigg] ,  \\
						{\cal{L}}_3^{ARR} &=&0.
	\end{eqnarray}
	
	Co-factors suffixed ${BRR}$; 
\begin{eqnarray}
		{\cal{L}}_1^{BRR}& =& 8M_\Lambda^2 \bigg[ -2 \gamma_\Lambda^2 \bigg]   , \\
				{\cal{L}}_2^{BRR}& =& 8M_\Lambda^2 \bigg[ 1\bigg] ,\\
						{\cal{L}}_3^{BRR}& =&0.
	\end{eqnarray}

Co-factors suffixed ${ARS}$  and ${ASR}$;
\begin{eqnarray}
	{\cal L}^{ARS}_4 &=& 2 M_{\Lambda }p_g (2\epsilon  \omega)^2 
	  \frac{1}{\gamma_\Lambda} \bigg[ \Phi_g \bigg] ,\\
{\cal L}^{ASR}_4 &=& 2 M_{\Lambda }p_g (2\epsilon  \omega)^2 
	  \frac{1}{\gamma_\Lambda} \bigg[ \Phi_h\bigg] .
	\end{eqnarray}
		
Co-factors suffixed ${ASS}$ ;
\begin{eqnarray}
	{\cal L}^{ASS}_1 &=&  ( p_{\Lambda } p_g 2\epsilon  \omega)^2 Z
	\bigg[ \Phi_{\bot}(X_a-2X_b)\nonumber \\
	&+&2 (L_0 +ZL_M/\gamma_\Lambda)
	\bigg],\label{KSS1}\\
	{\cal L}^{ASS}_2 &= &  (p_{\Lambda } p_g 2\epsilon  \omega)^2 Z
		\bigg[ \Phi_f X_a  \frac{1}{\gamma_\Lambda^2} \bigg],
		\label{KsS2}\\
	{\cal L}^{ASS}_3 &= &  (p_{\Lambda } p_g 2\epsilon  \omega)^2Z
	\bigg[ -Z\gamma_\Lambda L_M \bigg] .\label{KSS3}
\end{eqnarray}

Co-factors suffixed ${BSS}$;
\begin{eqnarray}
	{\cal L}^{BSS}_1 &=&  2(2p_{\Lambda }^2 p_g)^2 Z^2
	\bigg[ - 2X_b	\gamma_\Lambda^2\bigg],\label{JSS1}\\
	{\cal L}^{BSS}_2 &= &  2(2p_{\Lambda }^2 p_g)^2 Z^2
	\bigg[  X_a \bigg] ,\label{JSS2}\\
	{\cal L}^{BSS}_3 &= &  0.\label{JSS3}
\end{eqnarray}

The $X$, $L$, and, $\Phi$ functions are 
essential in building  the co-factors. They  are defined in  sect.\ VI. 
Important parameters are; the common lepton energies $\epsilon$ in the overall 
c.m.; the common hyperon energies $E_\Lambda$ in the hyperon c.m.; 
and $\omega$ the energy of the ISR photon. Three important relations 
between parameters are,
\begin{eqnarray}
\epsilon \omega&=&\epsilon^2-E_\Lambda^2, \\
Z&=&\frac{4 M_\Lambda^2}{Q^2}=1-\frac{1}{v_\Lambda^2},\\
\gamma_\Lambda&=& E_\Lambda/M_\Lambda, 
\end{eqnarray}  
with $Q^2=(p_1-p_2)^2$, and $v_\Lambda=p_\Lambda/E_\Lambda$ the 
hyperon velocity in the $\Lambda\bar{\Lambda}$ rest frame.

%
%
\newpage

\section{Angular variables}

The co-factors of sect.\ 4 are expressed in terms of functions 
$X$, $L$, and $\Phi$, which themselves  
are scalar functions of a set of unit vectors that we 
now proceed to define.

$\hat{\mathbf{k}}$ is a unit vector in the direction of motion
of the initial-state lepton in the overall cms;

$\mathbf{n}=\hat{\mathbf{q}}$ is  a unit vector in the direction of motion 
of the photon in the overall cms;

$\mathbf{g}(\mathbf{h})$ is  a unit vector in the direction of motion 
of the proton (antiproton) in the  rest system of the
$\Lambda (\bar{\Lambda})$ hyperon;

$\mathbf{f}(-\mathbf{f})$ is  a unit vector in the direction of motion 
of the $\Lambda (\bar{\Lambda})$ in the 
rest system of the hyperon pair; and finally

  $\mathbf{N}$ is a vector defined by,
\begin{equation}
\mathbf{N}=\frac{1}{ v\gamma} \bigg[\hat{\mathbf{k}}+(\gamma -1)
(\mathbf{n}\cdot\hat{\mathbf{k}})\mathbf{n}\bigg], \label{DefN}
\end{equation}
given the velocity $v$ and the function $\gamma(v)$ 
\begin{eqnarray}
v&=& \frac{\omega}{\sqrt{\omega^2+W^2}},\\
\gamma(v)&=& \frac{\sqrt{\omega^2+W^2}}{W}.
\end{eqnarray}
\newpage
\section{Angular functions}

The  $\Phi$	functions are by definition functions of 
 $\mathbf{n}$ and $\mathbf{N}$; 
\begin{eqnarray}
		\Phi_f&=&
		 (\mathbf{n}  \cdot\mathbf{f})^2
		+ ( \mathbf{N} \cdot\mathbf{f})^2, \\
		\Phi_{\bot}&=& (\mathbf{n}  \times\mathbf{f})^2
		+ ( \mathbf{N} \times\mathbf{f})^2,\\
	 \Phi_g &=& 
	       \mathbf{f}\cdot \mathbf{n} 
	\mathbf{f} \cdot ( \mathbf{g} \times \mathbf{n} ) +
	\mathbf{f}\cdot \mathbf{N}\mathbf{f} \cdot ( \mathbf{g} \times \mathbf{N} )
	 , \qquad\label{OMG}\\
	  \Phi_h  &=& 
  	 \mathbf{f}\cdot \mathbf{n} 
	\mathbf{f} \cdot ( \mathbf{h} \times \mathbf{n} ) +
	\mathbf{f}\cdot \mathbf{N}\mathbf{f} \cdot
	( \mathbf{h} \times \mathbf{N} ) ,\qquad  \label{OMH}\\
	\Phi &=& \mathbf{n}^2 +\mathbf{N}^2=
		\frac{1}{v^2}+ (  \mathbf{n} \cdot \hat{\mathbf{k}})^2 \nonumber\\
		&=&\Phi_f +
		\Phi_{\bot}.
\end{eqnarray}
 The  $\mathbf{N}$ vector is defined in  eq.(\ref{DefN}).
	%
%
	Introduced are also functions $L_0$ and $L_M$,
	\begin{eqnarray}
		L_0 &=&\mathbf{n}\cdot\mathbf{g}_\bot \,  \mathbf{n}\cdot \mathbf{h}_\bot +
		\mathbf{N} \cdot
		\mathbf{g}_\bot \, \mathbf{N} \cdot 
		\mathbf{h}_\bot ,\\
		L_M &=& (\mathbf{f}\cdot\mathbf{g} \mathbf{h}_\bot +
			\mathbf{f}\cdot\mathbf{h} \mathbf{g}_\bot )\nonumber \\
		&&\times
			( \mathbf{n}\mathbf{f}\cdot \mathbf{n} +
			\mathbf{N} \mathbf{f}\cdot\mathbf{N}  ),
	\end{eqnarray} 
	and the bilinear functions $X_a$ and $X_b$
\begin{eqnarray}
	X_a&=& 2 \mathbf{g} \cdot \mathbf{f} \mathbf{h} \cdot \mathbf{f} - 
	 \mathbf{g} \cdot \mathbf{h},\\
	  X_b&=& \mathbf{g} \cdot \mathbf{f} \mathbf{h} \cdot \mathbf{f} . 
	\end{eqnarray} 
	Here, orthogonal means orthogonal with respect to the $\mathbf{f}$ vector, i.e.,
	$\mathbf{g}_\bot=	\mathbf{g}	-\mathbf{f}(	\mathbf{f}\cdot	\mathbf{g}).$

\newpage
 \section{Phase space}

Since the intermediate-state hyperons represent states  whose masses 
may be considered fixed, 
it is useful to rewrite the phase-space differential  making this fact explicit,  by using the
 nesting formula \cite{Pil}
\begin{eqnarray}
	 \textrm{dLips}(k_1+k_2;q,l_1,l_2,q_1,q_2)&=& \nonumber\\
	\frac{1}{(2\pi)^2} \rd s_g \rd s_h
		\textrm{dLips}(k_1+k_2;q,p_1,p_2) && \nonumber \\	
	\times\textrm{dLips}(s_g;l_1,q_1)\, \textrm{dLips}(s_h;l_2,q_2),&&,
		 \label{Ph_space_def}
\end{eqnarray}
where by definition
\begin{eqnarray}
      p_1&=& l_1+q_1 ,  \\
			p_2 &=& l_2+q_2,
\end{eqnarray}
with $p_1^2=s_g$  and $p_2^2=s_h$.

Now, we know the analytic expressions for the two- and three-body phase-space
 differentials of eq.(\ref{Ph_space_def}) as given in ref.\cite{Pil}. For the two-body differential 
\begin{equation} 
\textrm{dLips}(s_g;l_1,q_1)
    = \frac{p_g\rd\Omega_g}{16\pi^2\sqrt{s_g}},
\end{equation}
where index $g$ reminds us we are in the cms of the 
 $N\pi$ pair, where each particle has momentum
 \begin{equation}
  p_g=\frac{1}{2\sqrt{s_g}}  \sqrt{\lambda(s_g,m_N^2,m_\pi^2)}.
 \end{equation}
 A corresponding differential, but with index $h$, refers 
to the contribution from the $\bar{N}\bar{\pi}$ pair.

The three-body differential of eq.(\ref{Ph_space_def}) is calculated after having been
reduced to a product of two-body differentials,
\begin{eqnarray}
	 \textrm{dLips}(s;q,p_1,p_2)&=& \nonumber\\
	\frac{1}{2\pi}{\rd s_c} \textrm{dLips}(s;p_c,q)	
	\textrm{dLips}(s_c;p_1,p_2),&&
		 \label{Ph_space_form}
\end{eqnarray}
with $p_c=p_1+p_2$ and $s_c=p_c^2$. Furthermore,  
\begin{equation}
s_c= s-2\omega\sqrt{s}, 
\end{equation}
so that $\rd s_c=-2\sqrt{s}\, \rd \omega$.

A straightforward evaluation  of the expressions for the two-particle differentials leads to
\begin{eqnarray}
   \textrm{dLips}(s;p_c,q) &=& \frac{\left| \mathbf{q}\right|\rd\Omega_q}{16\pi^2\sqrt{s}}, \\
\textrm{dLips}(s_c;p_1, p_2)
   & =& \frac{p_d\rd\Omega_d}{16\pi^2\sqrt{s_c}}.
\end{eqnarray}
Consequently, the cms momentum in each of these two-body cases equals
\begin{eqnarray}
p_q &=& \frac{1}{2\sqrt{s}}\sqrt{\lambda(s,s_c,0)}=\left|\mathbf{q}\right|, \\
 p_d&=&\frac{1}{2\sqrt{s_c}}  \sqrt{\lambda(s_c,s_g,s_h)}.
\end{eqnarray}

There are four two-body states to be reckoned with. 
They are characterized  by the following energies and momenta:
for the energies   
\[ \begin{array}{rlrl}
\sqrt{s}&= 2\epsilon & \sqrt{s_c}= 2E_\Lambda & \\
\sqrt{s_g}&= M_\Lambda \qquad &  \sqrt{s_h}= M_\Lambda, 
\end{array} \]
and for the momenta
\[ \begin{array}{rlrl}
 p_g&= p_\lambda  &
 p_q&= \|\mathbf{q} \|                  \\
  p_h&=  p_\lambda    & \qquad
 p_d&= p_\Lambda. 
\end{array} \]
with $p_\lambda$ defined as  the
\begin{equation}
p_\lambda = \frac{1}{2M_\Lambda} 
\sqrt{\lambda(M_\Lambda^2,m_N^2, m_\pi^2)}.
\end{equation}

Collecting all the information on   the phase-space differential 
of eq.(\ref{Ph_space_def}) gives,
\begin{align}
	&\textrm{dLips}(k_1+k_2;q,l_1,l_2,q_1,q_2)=       \nonumber\\
	& \frac{1}{(2\pi)^3} \rd s_c \rd s_g \rd s_h  \, \cdot
	\frac{p_q\rd \Omega_q}{(4\pi)^2 \sqrt{s}} \, \times \nonumber \\
&	 \frac{p_g\rd \Omega_g}{(4\pi)^2 \sqrt{s_g}} 	
	      		 \frac{p_h\rd \Omega_h}{(4\pi)^2 \sqrt{s_h}} 
	         \frac{p_d\rd \Omega_d}{(4\pi)^2 \sqrt{s_c}}.
	        			\label{Lipsg}
	\end{align}


We end this section with a remark. In the narrow 
width approximation 
the $\cal{K}$ factor of eq.(\ref{KM-factor}) contains 
two delta-function factors, 
 \begin{equation}
\delta(s_g-M_\Lambda^2)\delta(s_h-M_\Lambda^2),
\end{equation}
which cancel against the $\rd s_g$
and $\rd s_h$ integrations of eq.(\ref{Lipsg}).
The delta functions arise from  the identity 
\begin{equation}
\frac{1}{(s-M^2)^2+M^2\Gamma^2(\sqrt{s})} =\frac{\pi} { M\Gamma(M)}
\delta(s-M^2). \nonumber
\end{equation}

\newpage 
\section{Discussion}

In this paper we describe and analyze  the structure of the differential-cross-section 
 distribution in $e^+e^-$ annihilation for a $\Lambda\bar{\Lambda}$
final state, which is accompanied by an initial-state-radiation gamma. 
 To be specific, 
the final state is fixed by  the directional angles in the 
decays
$\Lambda \rightarrow N\pi$ and $\bar{\Lambda} \rightarrow \bar{N}\pi$, 
denoted $\Omega_g$ and $\Omega_h$, and the directional angle $\Omega_d$ 
in the  $\Lambda\bar{\Lambda}$
final state. Finally, there is the
gamma three-momentum-directional angle $\Omega_\omega$, and the radiated gamma energy $\omega$. 

We  start by rewriting 
the cross-section-distribution 
function  $\rd \sigma$
of eq.(\ref{dsigma}) as a product of three factors.
The first factor is
\begin{equation}
{\cal{K}}=\frac{1}{2s}\frac{(4\pi\alpha)^3}{(P^2)^2} 
\Bigg[\frac{\pi}{M_\Lambda \Gamma_\Lambda(M_\Lambda) }\Bigg]^2.
\end{equation} 
This is what remains when the  delta functions
$\delta(s_g-M_\Lambda^2)\delta(s_h-M_\Lambda^2)$ hidden
in the ${\cal{K}}$ factor of eq.(\ref{KM-factor}) are removed.

The second factor is the phase-space-density-distribution 
function. After having  absorbed the abovementioned delta functions
this factor reads
\begin{eqnarray}
	 \textrm{dLips}&=&
	 \frac{1}{(2\pi)^3}\  4\epsilon \rd \omega  
		\frac{\omega \rd \Omega_\omega}{(4\pi)^2 2\epsilon} 
		 \frac{p_\Lambda \rd \Omega_d}{(4\pi)^2 2E_\Lambda} 
		  \notag                   \\
  & \times &	\frac{p_\lambda \rd \Omega_g}{(4\pi)^2 2M_\Lambda} 
	 \frac{p_\lambda \rd \Omega_h}{(4\pi)^2 2M_\Lambda} , \label{lips}
\end{eqnarray}
where $p_\lambda = \sqrt{\lambda(M_\Lambda^2, m_N^2, m_\pi^2)/4M_\Lambda^2} $.

The third factor in the cross-section-distribution function is the squared matrix element 
$\overline{|{\cal{M}}_{red}|^2}$ of eq.(\ref{MM-decomp}),
to which we now turn our attention.

We expand 
$\overline{|{\cal{M}}_{red}|^2}$ on a set of functions, 
$A^{XY}$ and    $B^{XY}$,   which are bilinear forms  
in the  form factors  $G_M$ and $G_E$,
\begin{eqnarray}
A^{XY}(G_M,G_E) &=&|G_M|^2 {\cal{L}}^{AXY}_1 + |G_E|^2 {\cal{L}}^{AXY}_2
\nonumber \\
	  &+& 2 \Re (G_MG_E^{\star}) {\cal{L}}^{AXY}_3
		\nonumber \\ & + &2\Im (G_MG_E^{\star}) {\cal{L}}^{AXY}_4,
				\label{CDdef}
\end{eqnarray}
and similarly for $B^{XY}$. We refer to the set of  functions 
\{${\cal{L}}$\} as co-factors. Each of the suffixes $X$ and $Y$ 
stand for $R$ or $S$, where  suffix $R$ represents parity-conserving 
hyperon decay, and $S$ parity-violating hyperon decay.  
Several co-factors vanish; ${\cal{L}}^{ARR}_3={\cal{L}}^{BRR}_3=
 {\cal{L}}^{BSS}_3=0$. 

Co-factors suffixed $RR$ are independent of the hyperon decay angles 
$\Omega_g$ and $\Omega_h$; co-factors $RS$ and $SR$ are linear in one of 
the decay angles $\Omega_g$ or $\Omega_h$;  co-factors $SS$ 
are linear in both
$\Omega_g$ and $\Omega_h$.

	Until now, we have been concerned with the 
	$\gamma(N\pi)(\bar{N}\bar{\pi})$ final state, 
	 but it is also possible   to determine   
	the co-factors when one or both hyperons 
	remain intact, i.e., after integration  over 
   $\rd \Omega_g$ or  $\rd \Omega_h$ or both.

	 It turns out the co-factors 
	of sect.\ IV either vanish or remain intact. Thus, 
	
	Operation: $\int \rd \Omega_g\int \rd \Omega_h/(16\pi^2)$

	Co-factors; ${\cal L}^{ARR}_1 , {\cal L}^{ARR}_2 , 
{\cal L}^{BRR}_1 , {\cal L}^{BRR}_2 $.

Operation: $\int  \rd \Omega_g/(4\pi)$

	Co-factors; ${\cal L}^{ARS}_4.$ 
	
	Operation: $\int  \rd \Omega_h/(4\pi)$

	Co-factors; ${\cal L}^{ASR}_4.$ 
	
	Co-factors not listed vanish.
	
	\newpage

 
%
\newpage
\subsection*{Appendix}
The energy-momentum parameters describing the decay of Lambda  into  proton and  pion are,  
\begin{align}
	p_g&=\frac{1}{2M_\Lambda} \sqrt{\lambda(M_\Lambda^2, m_N^2, m_\pi^2)}, \\
	E_g  &=\frac{1}{2M_\Lambda}\big(M_\Lambda^2 +m_N^2-m_\pi^2\big) ,
\end{align}
representing the proton in the Lambda rest system. 

The weight functions $a_y$ and $b_y$ of eq.(\ref{MXY}), 
can be written as
\begin{eqnarray} 
 a_y &=& \frac{2}{(\epsilon \omega)^2\sin^2\theta} \Big[
  2E_\Lambda^2 \Big], \\
   b_y &=&\frac{2}{(\epsilon \omega)^2\sin^2\theta} \Big[
    2( \epsilon^4+ E_\Lambda^4) \nonumber \\
    &-&(\epsilon \omega)^2
   \sin ^2\theta\Big].
\end{eqnarray}
  We notice that $a_y$ and $b_y$ have different dimensions.


\newpage
\begin{thebibliography}{99}
\bibitem{BaBar} B. Aubert {\itshape et al.}~({\slshape BABAR} Collaboration),  Phys.~Rev.~D {\bf 76}, 092006 (2007)
\bibitem{Novo} L.\ V.\ Kardapoltzev, Bachelor's thesis, Novosibirsk 
              State University, 2007 (unpublished)
\bibitem{Czyz} H.\ Czy\.z, A. Grzeli\'nska, and J.\ H.\ K\"uhn, 
      Phys.~Rev.~D {\bf 75}, 074026 (2007)
	\bibitem{Ent1}
	 G\"oran F\"aldt, 
	Eur.\ Phys.\ J.\ {\bf A51}, 74  (2015) 
	\bibitem{EPJ58}
	 G\"oran F\"aldt, 
	Eur.\ Phys.\ J.\ {\bf A58}, 210  (2022)
  \bibitem{Pil} H.\ Pilkuhn, \textit{Relativistic Particle Physics}  
	(Springer-Verlag, Berlin, 1979)
			\bibitem{Eul} G\"oran F\"aldt, 
			Phys.~Rev.~D {\bf 97}, 053002 (2018)
			
\end{thebibliography}
 \end{document}